\documentclass[pre]{revtex4}
\usepackage{amsmath}
\usepackage{graphicx}
\begin{document}
\title{Damping of macroscopic oscillation and interference pattern in coupled Gross-Pitaevskii equations without self-interaction}
\author{Hidetsugu Sakaguchi and Fumihide Hirano}
\affiliation{Department of Applied Science for Electronics and Materials,
Interdisciplinary Graduate School of Engineering Sciences, Kyushu
University, Kasuga, Fukuoka 816-8580, Japan}
\begin{abstract}
Quantum phenomena appear in a macroscopic scale in Bose-Einstein condensates. The Gross-Pitaevskii (GP) equation describes the dynamics of the weakly interacting Bose-Einstein condensates. The GP equation has a form of the Schr\"odinger equation  with self-interaction. The coupled Gross-Pitaevskii equations are  used to describe some mixtures of Bose-Einstein condensates. In this paper, we will show some numerical results of coupled Gross-Pitaevskii equations without self-interaction, which has a form of nonlinearly-coupled Schr\"odinger equations. We demonstrate that the macroscopic oscillation and the interference of two quantum wave packets decay in time owing to mutual interaction, which is analogous to the decoherence in quantum mechanics of many particles.
\end{abstract} 
\maketitle
\section{Introduction}
Bose-Einstein condensates can exhibit the dynamics of wave function in a macroscopic scale. The Gross-Pitaevskii (GP) equation or the nonlinear Schr\"odinger equation is used to describe the dynamics of the weakly interacting Bose-Einstein condensates. In the GP equation, the mutual interaction among quantum particles is expressed through the average density, which is a kind of mean-field approximation~\cite{PS}. The GP equation is applied to many other physical systems such as optical fibers~\cite{KA}.  Solitons are typical nonlinear solutions to the GP equation and found in an experiment of Bose-Einstein condensates.~\cite{Kh}  The localized state appears as a balance between the dispersion effect of the quantum particle and  attractive interaction. 
Coupled Gross-Pitaevskii (CGP) equations are used to describe mutually interacting different types of quantum particles in Bose-Einstein condensates. The motion of solitons and the symmetry-breaking in the GP and CGP equations of two modes were studied by several authors.~\cite{Raghavan, Qian, Sakaguchi2} The CGP equations with two variables were often studied, but the model can be generalized to multi-mode equations, which can be applied for multi-mode optical fibers~\cite{CD}.  

Macroscopic quantum phenomena appear also in other systems such as Josephson junctions.~\cite{Jose} The quantum coherence can be destroyed by the interaction with the environment.~\cite{Zurek,Schlosshauer} The decoherence is mainly composed of dephasing and damping of macroscopic quantum oscillation. Caldeira and Leggett studied a theoretical model of a quantum harmonic oscillator coupled with a lot of small quantum harmonic oscillators and showed that the macroscopic quantum oscillation decays owing to the effective friction with the surrounding quantum particles~\cite{CL, Leggett, Fujikawa}. 

In this paper, we study the CGP equations without self-interaction terms. In the model equations, each wave function obeys the Schr\"odinger equation if mutual interaction terms are neglected, and the mutual interaction is expressed with the nonlinear terms. That is, our model has a form of nonlinearly-coupled Schr\"odinger equations. The CGP equations are obtained by assuming a mean-field type approximation, where the coupling coefficient is proportional to the s-wave scattering length between the different types of particles. We demonstrate the damping of macroscopic oscillation with the numerical simulation of the CGP equations in harmonic and double-well potentials.
   
\section{Quantum Mechanics and Variational Approximation}
The one-dimensional Schr\"odinger equation in a potential $U(x)$ is expressed as
\begin{equation}
i\frac{\partial \phi}{\partial t}=-\frac{1}{2m}\frac{\partial^2\phi}{\partial x^2}+U(x)\phi \label{SH},
\end{equation}
where $\phi$ denotes the wave function, $m$ is the mass, and the Planck constant $\hbar$ is set to 1. 
The Schr\"odinger equation can be solved for $U(x)=0$.  If the initial condition is expressed with the Gaussian function:
\[\phi(x,0)=\left (\frac{2}{\pi \sigma^2}\right )^{1/4}\exp(-x^2/\sigma^2),\] 
$\phi(x,t)$ at time $t$ is expressed as
\begin{equation}
\phi(x,t)=\left (\frac{2}{\pi\sigma^2(t)}\right )^{1/4}\exp\{-x^2/\sigma^2(t)+i\beta(t)x^2\}, \label{wf2}
\end{equation}
where \[\sigma^2(t)=\sigma_0^2+\frac{4t^2}{m^2\sigma_0^2},\; \beta(t)=\frac{2t}{m\sigma_0^2\sigma^2(t)}.\]
The width of the wave packet increases with $t$. This is the dispersion effect of the quantum mechanics. As the width $\sigma_0$ of the initial wave packet is smaller, the width $\sigma(t)$ increases more rapidly. 

In the harmonic potential $U(x)=(1/2)kx^2$, the wave function in the ground state is expressed as 
\begin{equation}
\phi=Ae^{-\alpha x^2-i\mu t}, \label{HO}
\end{equation}
where $\alpha=\sqrt{mk}/2$ and the norm is given by $N=\int |\Phi|^2dx=A^2\sqrt{\pi/(2\alpha)}$. The Gaussian wave packet exhibits the simple oscillation in the harmonic potential. That is, the wave function:  
\begin{equation}
\phi=Ae^{-\alpha (x-\xi(t))^2+ip(x-\xi(t))-i\mu t}, \label{wf3}
\end{equation}
where $\xi(t)=X\cos(\omega t)$ and $p=-mX\omega \sin(\omega t)$ with $\omega=\sqrt{k/m}$ is an exact solution to Eq.~(\ref{SH}). The center of mass of the wave function exhibits the simple oscillation $\xi(t)=X\cos(\omega t)$. The frequency is the same as the one in the classical mechanics. 

In the variational approximation (VA), the wave function is assumed to be 
\begin{equation}
\phi=A\exp\{-\alpha (x-\xi(t))^2+i\beta (x-\xi(t))^2+ip (x-\xi(t))+i\mu t\}. \label{wf}
\end{equation}
The Lagrangian for the Schr\"odinger equation is written as 
\begin{equation}
L=\int \{i(\phi_t\phi^{*}-\phi_t^{*}\phi)-1/(2m)|\partial \phi/\partial x|^2-(1/2)kx^2|\phi|^2\}dx.
\end{equation}
Substitution of Eq.~(\ref{wf}) into the Lagrangian yields
\begin{equation}
L_{eff}=N\{-\beta_t/(4\alpha)-(1/2m)(\alpha^2+\beta^2)/\alpha+p\xi_t-p^2/(2m)-(1/2)k\xi^2\}.
\end{equation}
The Euler-Lagrange equation for the effective Lagrangian $L_{eff}$ is expressed as 
\[\frac{d}{dt}\frac{\partial L_{eff}}{\partial z_t}=\frac{\partial L_{eff}}{\partial z}.\] Four variables $z=\alpha, \beta,\xi$, and $p$ obey  
\begin{equation}
\frac{d\alpha}{dt}=-\frac{4\alpha\beta}{m},\; \frac{d\beta}{dt}=\frac{2}{m}(\alpha^2-\beta^2)-\frac{k}{2},\; \frac{d\xi}{dt}=\frac{p}{m},\;\frac{dp}{dt}=-k\xi. \label{va}
\end{equation}
If $k$ is set to be 0 or $U(x)=0$, 
\begin{equation}
\alpha=\frac{1}{\sigma_0^2+4t^2/(m^2\sigma_0^2)},\; \beta(t)=\frac{2t}{m\sigma_0^2\{\sigma_0^2+4t^2/(m^2\sigma_0^2)\}}
\end{equation}
are solutions to Eq.~(\ref{va}). 
The wave function Eq.~(\ref{wf2}) is exactly expressed by this VA method.
For $k\ne 0$, there is a solution, $\beta(t)=0$, $\alpha(t)=\sqrt{mk/2}$, $\xi(t)=X\cos(\omega t)$ with $\omega=\sqrt{k/m}$. The wave function Eq.~(\ref{wf3})
 is expressed exactly by this VA method. 
\section{Damping of Macroscopic Oscillation in Coupled Gross-Pitaevskii Equations}
In the classical Brownian motion, a macroscopic particle interacts with many small particles. The macroscopic motion decays and fluctuates in time. The Caldeira-Leggett theory predicts that  a similar process occurs even in quantum systems and the macroscopic quantum oscillation decays owing to the interaction with surrounding many quantum particles. However, it is very hard to perform direct numerical simulation of the $n$-particle Schr\"odinger equation for large $n$. We would like to demonstrate a similar process using the CGP equations. The CGP equations in an external potential $U(x)$ are expressed as   
\begin{eqnarray}
i\frac{\partial \Phi}{\partial t}&=&-\frac{1}{2M}\frac{\partial^2\Phi}{\partial x^2}+U(x)\Phi-\gamma\sum_{i=1}^n|\phi_i|^2\Phi, \label{GP3} \nonumber\\
i\frac{\partial \phi_i}{\partial t}&=&-\frac{1}{2m_i}\frac{\partial^2\phi_i}{\partial x^2}-\gamma|\Phi|^2\phi_i.  \label{GP30}
\end{eqnarray}
Here, the external potential $U(x)$ is applied only for $\Phi$.  
The inverse mass $1/M$ for $\Phi$ is set to be 0.2 and the inverse mass $1/m_i$ for $\phi_i$ is chosen from a uniform random number between 0 and 0.4.  The norm of $\Phi$ is $N_0=4$ and the  norm of $\phi_i$ is assumed to be $N_i=0.5$. In this section, we consider the case of the harmonic potential $U(x)=(1/2)kx^2$ with $k=0.25$. 
Figure 1 show the time evolution of the center of mass $X=\int x|\Phi|^2dx/N_0$ at (a) $\gamma=0.01$ and (b) 0.03 for $n=40$. The initial condition is $\Phi=(32\alpha_{0s}/\pi)^{1/4}\exp\{-\alpha_{0s}(x-x_0)^2\}$ and $\phi_1=\{\alpha_{is}/(2\pi)\}^{1/4}\exp(-\alpha_{is}x^2)$ where $x_0=0.5$ and $\alpha_{0s}$ and $\alpha_{is}$ are obtained by the VA approximation. The center of mass of $\Phi$ is shifted by $x_0$. At $\gamma=0$, the simple oscillation occurs as shown in Sec.2. Owing to the mutual interaction, a damping oscillation of $X$ appears.  As $\gamma$ is increased, the damping rate increases. The amplitude did not decay to zero but chaotic fluctuations are maintained because of the finiteness of $n$.    
\begin{figure}[h]
\begin{center}
\includegraphics[height=4.cm]{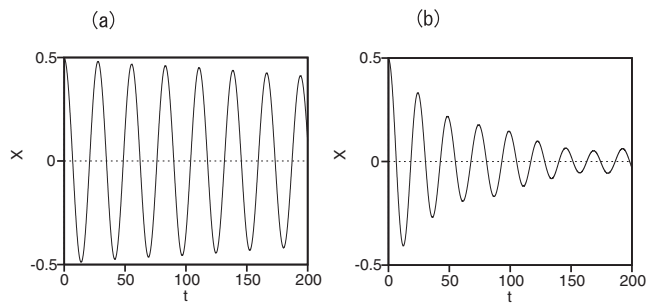}
\end{center}
\caption{Time evolutions of the center of mass $X$ at (a) $\gamma=0.01$ and (b) 0.03 by the CGP equations Eq.~(\ref{GP30}) with $n=40$ in a harmonic potential $U(x)=0.125x^2$.}
\label{fig1}
\end{figure}
\begin{figure}[h]
\centering
\includegraphics[height=3.5cm]{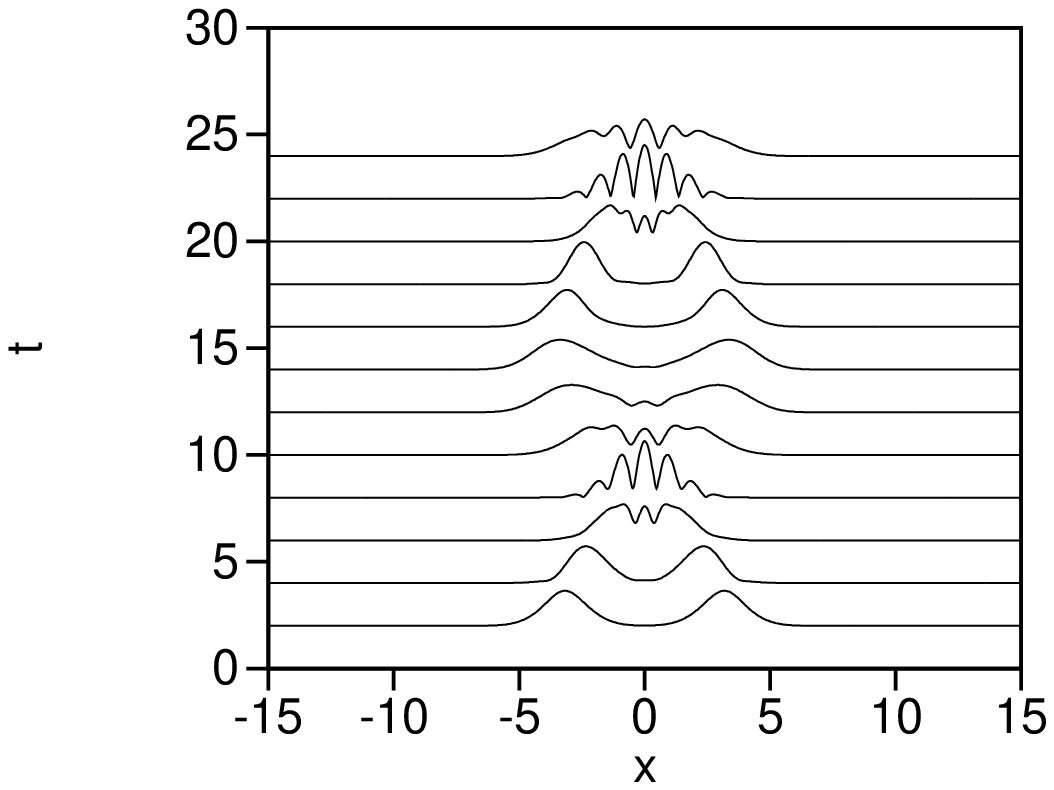}
\includegraphics[height=3.5cm]{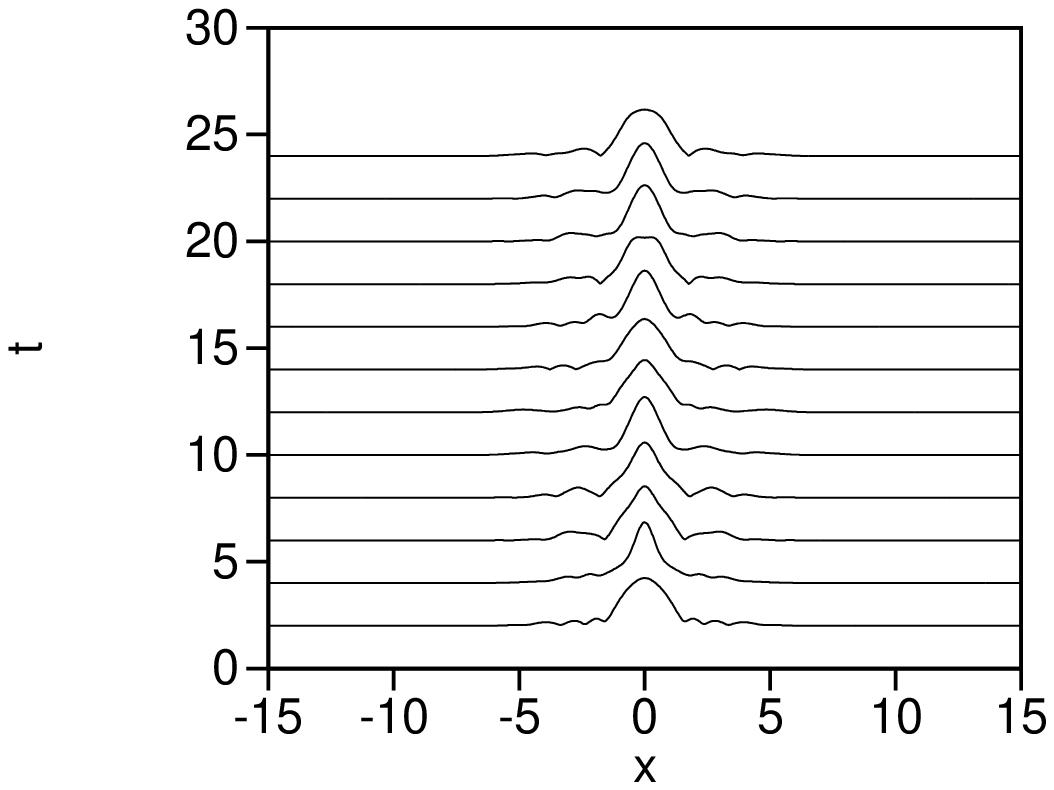}
\caption{Time evolution of $|\Phi|$ at $\gamma=0.05$ for $0<t<24$ (left) and $160\pi<t<160\pi+24$ (right). }
\label{fig2}
\end{figure}

The damping of macroscopic oscillation is related to the disappearance of macroscopic interference. We have performed a numerical simulation of Eq.~(\ref{GP30}). The parameters are the same as the ones in Fig.~1. 
The initial condition is set to be $\Phi=(32\alpha_{0s}/\pi)^{1/4}\{\exp\{-\alpha_{0s}(x-x_0)^2\}+\exp\{-\alpha_{0s}(x+x_0)^2\}$ and $\phi_1=\{\alpha_{is}/(2\pi)\}^{1/4}\{\exp(-\alpha_{is}(x-x_0)^2)+\exp(-\alpha_{is}(x-x_0)^2)\}$ with $x_0=3.5$. That is, two Gaussian wave packets separated by $2x_0$ are set initially. The width  $W=\{\int x^2 |\Phi|^2dx/\int |\Phi|^2dx\}^{1/2}$ exhibits a damping oscillation at $\gamma=0.05$, which suggests the damping of interference.  Figures 2(a) and (b) show the time evolutions of $|\Phi|$ at $\gamma=0.05$ (a) for $0<t<24$ and (b) $160\pi<t<160\pi+24$. In the initial stage, a standing wave pattern appears near $x=0$ at the collision of two wave packets, because of the interference of the two wave packets. However, the two wave packets merge and the interference pattern disappears as shown in Fig.~2(b).   

\section{Decay of Macroscopic Quantum Tunneling in Coupled Gross-Pitaevskii Equations.}
In this section, we consider the CGP equations (\ref{GP30}) with the double-well potential $U(x)=-(1/2)x^2+(1/8)x^4$. At first, we consider the Schr\"odinger equation (1) with the double-well potential. 
Figure 3(a) shows the ground state $\Phi_0$ (solid line) which is the mirror symmetric with respect to $x=0$, and the first excited state $\Phi_1$ (dashed line) which is anti-symmetric around $x=0$ for $M=5$.  The energies of the ground and excited states are respectively $E_0=-0.2334$ and $E_1=-0.1963$, and the energy difference is $\Delta E=0.0371$. 
If the initial condition is given as $\Phi(0)=(1/\sqrt{2})(\Phi_0+\Phi_1)$, $\Phi(t)$ is expressed as $\Phi(t)=(1/\sqrt{2})(\Phi_0e^{-iE_0 t}+\Phi_1e^{-iE_1t})$. The center of mass $X=\int x|\Phi(t)|^2dx/\int |\Phi|^2dx$ exhibits the oscillation with period $T=2\pi/\Delta E$. Figure 3(b) shows the time evolutions of $X$ obtained by direct numerical simulation (solid line) of Eq.~(\ref{SH}) and $X=1.18\cos(0.0371 t)$ (dashed line). 
The oscillation of period $T=2\pi/\Delta E$ is well reproduced in direct numerical simulation of Eq.~(\ref{SH}). 
 Even if the average energy $E=(E_0+E_1)/2$ is below the local maximum $U=0$ at $x=0$, the quantum particle exhibits the oscillation through the potential wall at $x=0$. This is the oscillation by the tunneling effect in the quantum mechanics and a kind of quantum coherence. 

\begin{figure}[h]
\begin{center}
\includegraphics[height=4.cm]{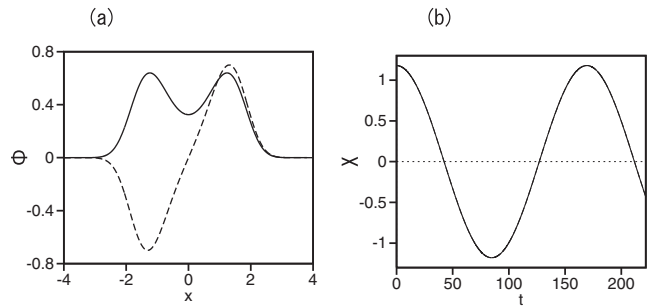}
\end{center}
\caption{(a) Wave function in the ground state (solid line) and the first excited state (dashed line) for the Schr\"odinger equation (1). (b) Time evolution of the center of mass $X$ (solid line)starting from $1/\sqrt{2}(\Phi_0+\Phi_1)$, and $X=1.18\cos(0.0371 t)$ (dashed line). The two lines overlap well.}
\label{fig3}
\end{figure}
\begin{figure}[h]
\begin{center}
\includegraphics[height=3.7cm]{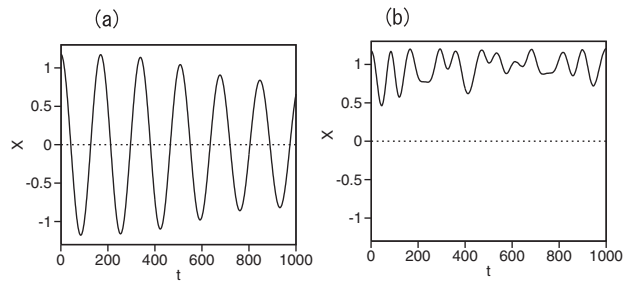}
\end{center}
\caption{Time evolutions of the center $X$ of mass for $\Phi$  at (a) $\gamma=0.005$ and (b) 0.04. The norms are $N_0=4$ and $N_i=1$, and $M=5$ and $1/m_i$ is a uniform random number between 0 and $0.4$.}
\label{fig4}
\end{figure}
\begin{figure}[h]
\begin{center}
\includegraphics[height=3.7cm]{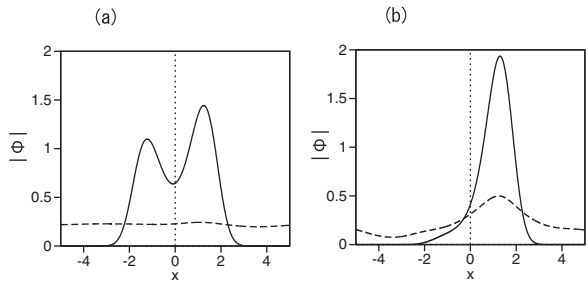}
\end{center}
\caption{Snapshot profiles of $|\Phi|$ (solid line) and $|\phi_2|$ (dashed line) at (a) $\gamma=0.0125$, (b) $0.04$ in the case that  $1/m_i$ is a uniform random number between 0 and $0.4$. }
\label{fig5}
\end{figure}
\begin{figure}[h]
\begin{center}
\includegraphics[height=3.7cm]{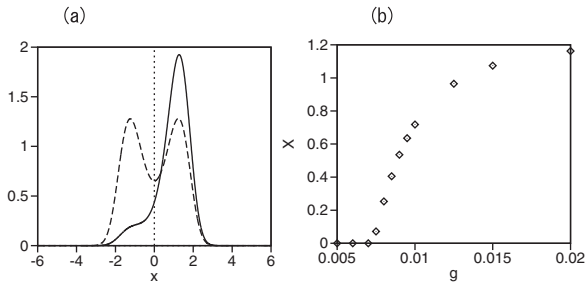}
\end{center}
\caption{(a) Profiles of $|\Phi|$ in the ground state at $\gamma=0.02$ (solid line) and 0.005 (dashed line) obtained by the imaginary-time evolution method.(b) The center of mass $X$ of the ground state as a function of $\gamma$.
 }
\label{fig6}
\end{figure}
We performed numerical simulation of CGP equations Eq.~(\ref{GP30}) with  $U(x)=-(1/2)x^2+(1/8)x^4$. The number $n$ is set to be 20. The initial conditions are $\Phi=\sqrt{2}(\Phi_0+\Phi_1)$ and $\phi_i=\sqrt{1/L_0}$ with $L_0=20$. The norms are $N_0=4$ and $N_i=1$, and $M=5$ and $1/m_i$ is a uniform random number between 0 and $0.4$.  
Figure 4 shows the time evolutions of the center $X$ of mass for $\Phi$  at (a) $\gamma=0.005$ and (b) 0.04. $X$ decays toward 0 at $\gamma=0.005$, however, fluctuates around 0 because $n$ is finite. The fluctuations  appear necessarily owing to the fluctuation-dissipation relation in thermally excited states. At $\gamma=0.04$, a symmetry-breaking occurs and $X$ fluctuates around 1. 
This is because the localization transition occurs around $\gamma=0.02$ and a soliton-like structure is formed for larger $\gamma$. Large fluctuations appear at $\gamma=0.02$ probably because $\gamma=0.02$ is close to the critical value.  
Figures 5(a) and (b) show snapshot profiles of $|\Phi|$ and $|\phi_2|$ with $1/m_2=0.242$ at $t=1000$ at $\gamma=0.0125$ and 0.04. At $\gamma=0.0125$, $|\Phi|$ takes a two-peak structure, which is close to the ground state of the double-well potential. This might be interpreted that the initial superposed state $\sqrt{2}(\Phi_0+\Phi_1)$ is destroyed and decays toward the ground state $2\Phi_0$. 
At $\gamma=0.04$, $|\Phi|$ is localized near $x=1.3$ and $\phi_2$ also have a peak near $x=1.3$. In the symmetry-broken localized state. the uncertainty of the position in the macroscopic state is reduced compared to the symmetric ground state of the Schr\"odinger equation, and $X$ is trapped in one minimum of the double well potential. This might be interpreted that the behavior of a classical particle with viscous force in a double-well potential is recovered.   

The ground state of the CGP equations can be calculated by the imaginary-time evolution of Eqs.~(\ref{GP30}) with the double-well potential. The wave function $|\Phi|$ is mirror-symmetric at $\gamma=0.005$ and the symmetry-breaking occurs and $|\Phi|$ is localized near $x=1.3$ at $\gamma=0.02$ owing to the attractive interaction.
Figure 6(a) shows the center of mass $X$ of the ground state as a function of $\gamma$. Figure 6(b) shows the center of mass $X$ of the ground state as a function of $\gamma$. The symmetry-breaking occurs at $\gamma\simeq 0.0075$ in the ground state. The symmetry-breaking occurs at higher $\gamma$ in direct numerical simulations starting from the initial condition $\Phi=\sqrt{2}(\Phi_0+\Phi_1)$ and $\phi_i=\sqrt{1/L_0}$ as shown in Fig.~4 because the total energy is higher than that of the ground state. 
  
\section{Summary}
We have performed a few numerical simulations of the coupled Gross-Pitaevskii equations without self-interaction to show the damping of macroscopic oscillation and interference pattern. Firstly, we have shown the validity of the VA method for the time evolution of the Schr\"odinger equation.  Then, we have demonstrated the damping of macroscopic quantum oscillation owing to the mutual interaction using the CGP equations in the harmonic potential. The disappearance of interference was also confirmed owing to mutual interaction in the CGP equations. Finally, we have demonstrated the damping of tunneling oscillation in a double-well potential. For small $\gamma$, $\Phi$ approaches the mirror-symmetric ground state of the Schr\"odinger equation. For large $\gamma$, $\Phi$ is localized near one well of the double-well potential, and the dynamics of the center of mass of the localized state is similar to the particle motion in the classical mechanics with dissipation.

\end{document}